# Non-destructive testing and evaluation of composite materials/structures: A state-of-the-art review

Bing Wang[1,2]*, Shuncong Zhong[2], Tung-Lik Lee[3], Kevin S Fancey[4], Jiawei Mi[4]

[1] Department of Engineering, University of Cambridge, Cambridge CB2 1PZ, UK
[2] School of Mechanical Engineering and Automation, Fuzhou University, Fuzhou 350116, China
[3] ISIS Neutron and Muon Source, Rutherford Appleton Laboratory, Harwell Oxford, Didcot, OX11 0QX, UK
[4] Department of Engineering, University of Hull, Hull, HU6 7RX, UK
* Corresponding to: bw407@cam.ac.uk (B Wang)

**Abstract:**

Composite materials/structures are advancing in product efficiency, cost-effectiveness and the development of superior specific properties. There are increasing demands in their applications to load-carrying structures in aerospace, wind turbines, transportation, and medical equipment, *etc.* Thus robust and reliable non-destructive testing (NDT) of composites is essential to reduce safety concerns and maintenance costs. There have been various NDT methods built upon different principles for quality assurance during the whole lifecycle of a composite product. This paper reviews the most established NDT techniques for detection and evaluation of defects/damage evolution in composites. These include acoustic emission, ultrasonic testing, infrared thermography, terahertz testing, shearography, digital image correlation, as well as X-ray and neutron imaging. For each NDT technique, we cover a brief historical background, principles, standard practices, equipment and facilities used for composite research. We also compare and discuss their benefits and limitations, and further summarise their capabilities and applications to composite structures. Each NDT technique has its own potential and rarely achieves a full-scale diagnosis of structural integrity. Future development of NDT techniques for composites will be directed towards intelligent and automated inspection systems with high accuracy and efficient data processing capabilities.

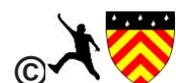

# LIST OF CONTENTS

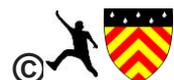

# LIST OF FIGURES

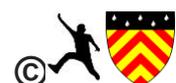

# LIST OF TABLES

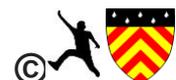

# 1    Introduction

Composite materials/structures are advancing in product efficiency, cost effectiveness, and the development of superior specific properties (strength and modulus). There are increasing demands in their applications to load-bearing structures in aerospace, wind turbines, transportation, and medical equipment *etc*. [1]. Manufacturing of composite materials is a multivariable task, involving many procedures, where various types of defects may occur within a composite product, giving rise to significant safety concerns in service [2]. Detection and evaluation to maintain structural integrity are particularly challenging since composites are usually non-homogeneous and anisotropic. Defects and damage can occur within numerous locations at various levels of scale, making it difficult to track all the damage sites which can result in complex damage mechanisms [3]. In addition, damage accumulation within a composite is closely related to the actual strength, stiffness and lifetime prediction of the component. Therefore, robust and reliable non-destructive testing (NDT) of composites is essential for reducing safety concerns, as well as maintenance costs [4] to minimise possibilities for process disruption and downtime. These factors attract interest both from academic researchers and industrial engineers.

There are a wide variety of NDT techniques built upon different principles. These have demonstrated effectiveness in quality assurance throughout the whole lifecycle of composite products; *i.e.* in process design and optimisation, process control, manufacture inspection, in-service detection, and structural health monitoring [5]. There are reviews available on NDT methods used for composites research over different timelines, focusing on various aspects: for general methods and trends over last 30 years refer to [4,6–11]; specific areas include those which concentrate on porosity in composite repairs [12], crack damage detection [13], bond defect determination in laminates [14], thick-wall composites [15], sandwich structures [16,17], large-scale composites [18], smart structures [19], as well as inspection and structural health monitoring of composites [20,21], especially for marine [22], wind turbine [23–26], and aerospace applications [27,28]. Audiences are recommended to refer to further information on their specific interests.

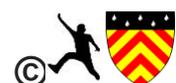

This paper reviews the most established NDT techniques for detection and evaluation to ensure the structural integrity of composite materials/structures; however, a full description of all methods is beyond the scope of this paper. Instead, we aim to provide a practical review of the established and emerging NDT techniques and their applications to composite research. The American Society for Testing and Materials (ASTM) has developed more than 130 standards, guides, and practices, containing technical specifications, criteria, requirements, procedures and practices for most of the NDT techniques [29]. We also include the standard practices for each NDT method available from the ASTM to provide guidance for researchers and engineers. These make it a unique state-of-the-art review paper to cover the most up-to-date practical information for NDT techniques and their applications to composite materials and associated structures.

The paper is organised as follows. Section 2 introduces the potential defects and damage evolution in composites. Section 3 provides an overview of development and principles of NDT techniques, and then elaborates on eight well-established NDT methods in subsections, covering a brief historic background, principles, standard practices, equipment and facilities for each NDT method in composite research. These include visual inspection (Section 3.1), acoustic emission (Section 3.2), ultrasonic testing (Section 3.3), infrared thermography (Section 3.4), terahertz testing (Section 3.5), shearography (Section 3.6), digital image correlation (Section 3.7), as well as X-ray and neutron imaging (Section 3.8). Section 4 compares and discusses the benefits and limitations of above NDT techniques, and further summarises their capabilities and applications to composite structures. This paper is concluded by the further development of NDT techniques, which is driven by intelligent and automated inspection systems with high accuracy and efficiency in data processing.

## 2      Defects and damage evolution in composites

Manufacturing-induced flaws/defects can occur in many forms: unevenly distributed fibres cause resin-rich regions; laminate-tool interactions result in in-plane fibre waviness or out-of-plane fibre wrinkling [30,31]; voids and porosities arise from poor resin infusion; inclusions from contaminated ambient conditions; misalignment of ply and fibre orientation; matrix cracking, laminate warping and buckling from build-up of thermal residual stresses during curing *etc.* [32,33]. Flaws/defects act as stress concentration points, promoting crack propagation and delamination to reduce effective strength, stiffness and service time of composite products [34]. Although residual stresses can occasionally be beneficial, especially for producing morphing composite structures [35–38], they are usually detrimental [32]. A wide range of processes have developed for the moulding of composite materials to reduce flaws, defects and build-up residual stresses that may occur during manufacture. These can involve multi-step processing, expensive consumables and equipment, to meet technical requirements. Typical industrial practice generally includes NDT inspection and evaluation of composite products to ensure their structural integrity and mechanical performance, which can be particularly challenging [39].

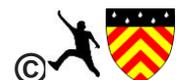

Figure 1 summarises the typical flaws and defects that may occur during manufacturing, and the in-service damage evolution of a composite material/structure. There are no clear boundaries on the scales of different defects and damage (which also depend on composite constituents); thus here we provide a general guidance according to the published literature.

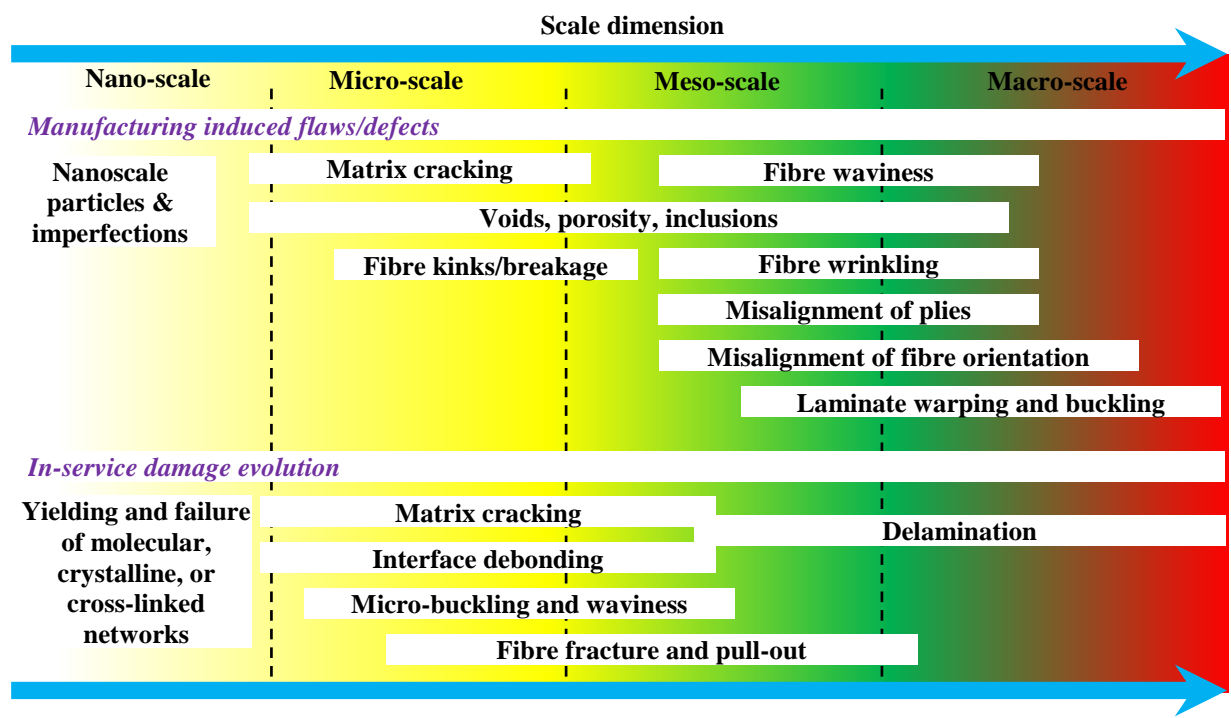

**Figure 1    Manufacturing-induced flaws/defects, and in-service damage evolution of a composite material/structure, with their potential scale dimensions.**

In-service damage evolution within a composite material/structure depends on composite constituents and loading conditions. Their failure processes are an accumulation of basic rupture mechanisms that include matrix cracking, fibre/matrix debonding, fibre fracture and pull-out, micro-buckling and waviness, delamination, *etc*. [40,41]. The damage process initiates at the nano- or micro-scale, where molecular chains, crystals and amorphous regions (for semi-crystalline thermoplastic polymers) or cross-linked molecular networks (for thermosetting polymers) carry loads until their limits are reached; damage then starts to accumulate on the micro-scale through crack propagation, interface debonding and micro-buckling, fibre fracture and pull-out, which lead to delamination, ultimately developing into macro-scale failure.

# 3    Non-destructive testing & evaluation techniques

The term 'non-destructive testing' covers a wide range of analytical techniques to inspect, test or evaluate chemical/physical properties of a material, component or system without causing damage. Early established NDT techniques include ultrasonic, X-ray radiography, liquid penetrant testing, magnetic-particle testing and eddy-current testing, which were initially developed for steel industry. Among these, ultrasonic and radiographic detection are also effective inspection techniques for composite structures [17]. It is difficult to select appropriate NDT techniques for a specific purpose; however, ASTM E2533 [5] serves as a practical guide in using NDT methods on composite materials/structures for aerospace applications.

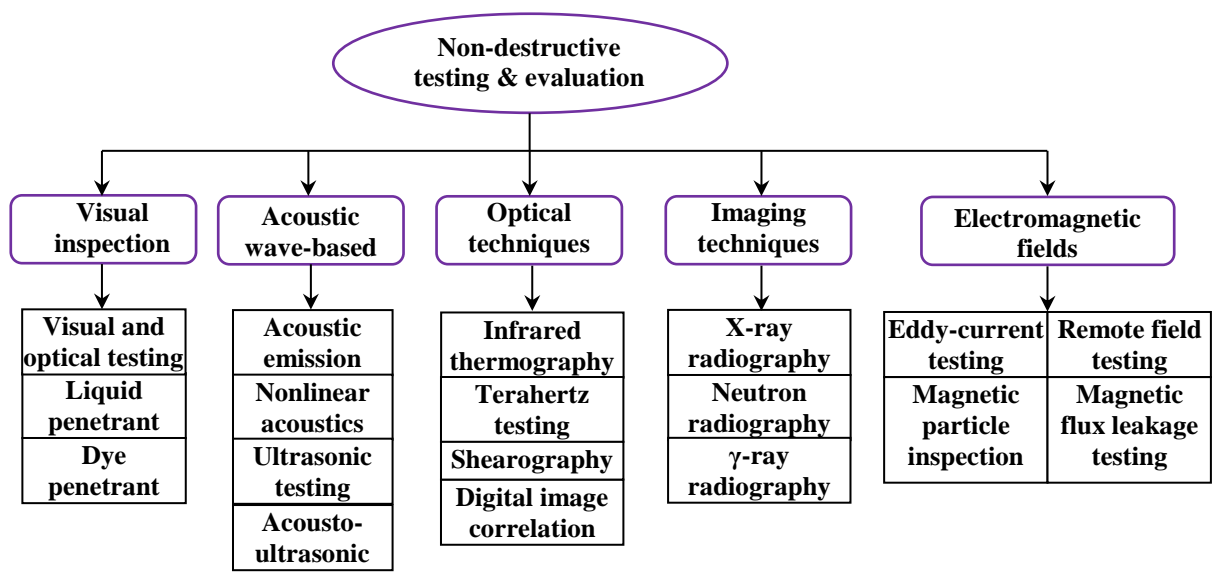

**Figure 2      Categories of non-destructive testing and evaluation techniques.**

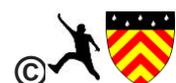

To date, there have been numerous NDT methods based on different principles, see Figure 2. They can be categorised into five groups: (i) visual inspection (i.e. those visible to the human eye); (ii) acoustic wave-based techniques, such as acoustic emission, nonlinear acoustics, and ultrasonic waves; (iii) optical techniques, which include infrared thermography, terahertz testing, shearography, digital image correlation; (iv) imaging-based techniques, *e.g.* X-ray/neutron radiography/tomography and micro-tomography [4]; (v) electromagnetic field based techniques, such as eddy-current testing, remote field testing, magnetic-particle inspection and magnetic flux leakage testing [42].

Here, we focus on eight established and emerging NDT techniques and their applications to composite research in categories (i) to (iv), with the exclusion of category (v). Since NDT methods in (v) are based on electromagnetic induction, their applications are limited to conductive materials [43]. Eddy-current testing (ECT) for example, is well-established and widely used for detecting cracks and corrosion in homogeneous metallic materials. Although it may be applicable to carbon composites, their conductivities are usually very low and inhomogeneous due to the lay-up and bundling of the conductive fibres [44]. This leads to further issues and difficulties for ECT to be an efficient and cost-effective solution for most composite NDT applications.

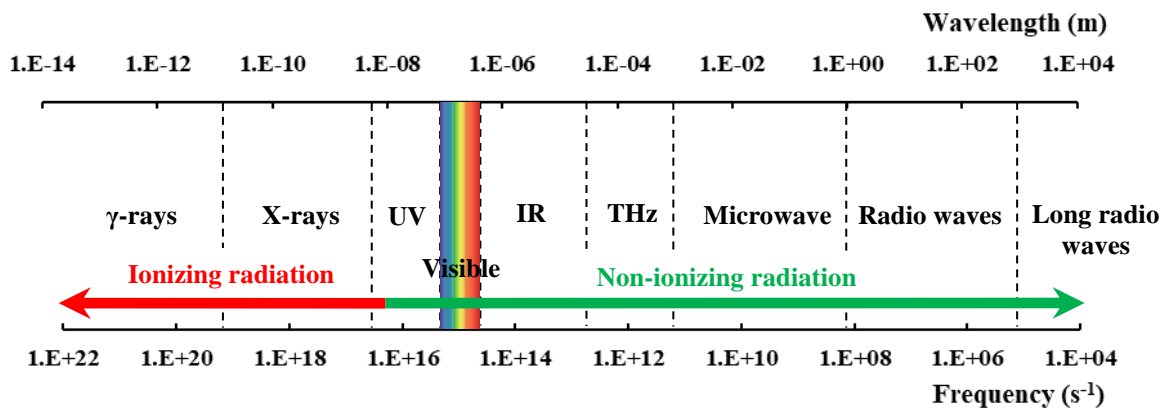

**Figure 3**     **Diagram of the electromagnetic spectrum, defining the various regions of radiation according to their range of frequencies and wavelengths.**

The measurement principles of each elaborated NDT technique depend on the characteristics of the electromagnetic waves based. Figure 3 shows the electromagnetic spectrum with divided wavelength sub-regions: the soft boundaries indicate terminologies for the subsections. Developments in generation and detection within each spectral regime have induced numerous industrial applications [45]. Ionizing radiation consists of short-wave ultraviolet (UV), X-rays, gamma-rays or highly energetic particles, such as α-particles, β-particles or neutrons, which are harmful to biological tissues; whereas the remaining part of the spectrum is considered to be non-ionizing radiation.

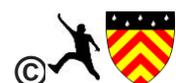

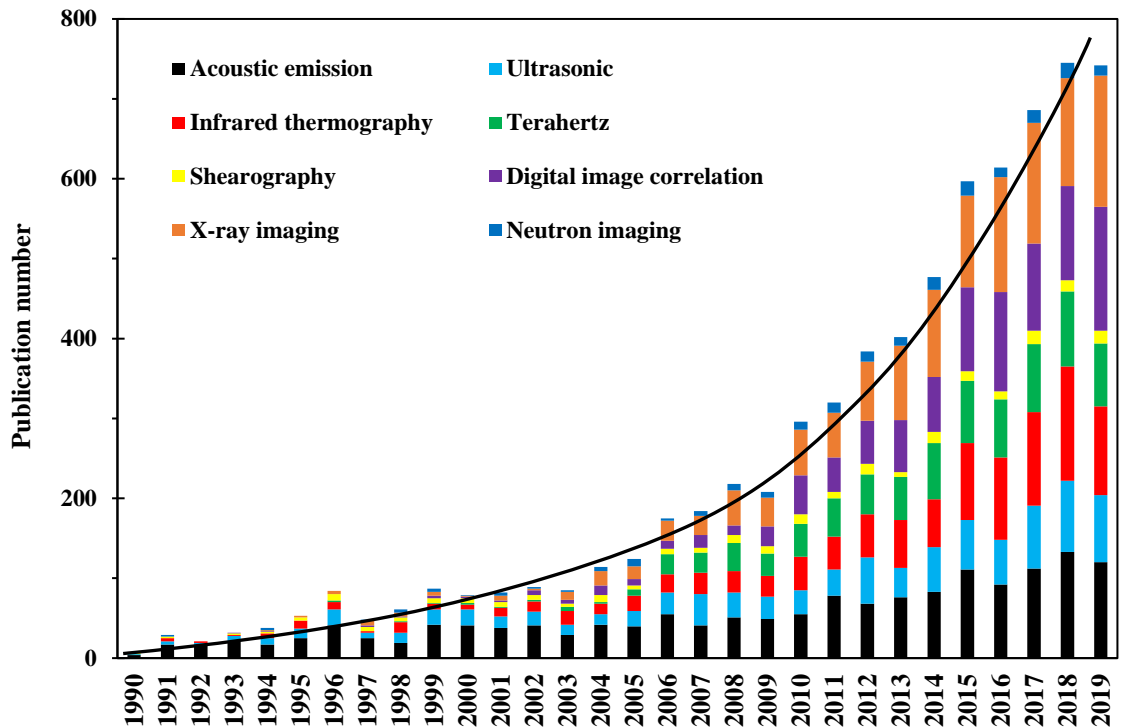

**Figure 4** **A comparison of publication numbers on various NDT methods and their applications to composite materials/structures in the last 30 years; data are retrieved from Web of Science Core Collection database.**

To date, there have been a growing number of research activities in this field. We performed an electronic data-base search on articles published in last 30 years (until 31/12/2019) using the Web of Science Core Collection database, to trace the trends in using various NDT techniques within composite research, see Figure 4. The use of acoustic emission on composites has a long history and is well-established; it is still active in a relatively steady state. Owing to significant developments in equipment manufacture, computing power, imaging processing and acquisition techniques over last three decades, there have been rapid increases in the application of infrared thermography, ultrasonic, digital image correlation and X-ray imaging to non-destructive detection and evaluation of composite materials/structures. Terahertz (THz)-based NDT technology has become a promising technique for composite inspection within last decade [46,47]. Research papers on the shearography technique is low, but it was promoted significantly by the invention of the laser in the 1960s [48], thus it is well-established and widely used for industrial NDT, especially in aerospace [49,50]. Although neutron imaging shares similar principles to X-ray imaging, the generation of neutrons is more expensive than for X-rays, the former requiring either a nuclear reactor or spallation process [51]. This has resulted in relatively few publications on its application to composite materials.

## 3.1    Visual inspection

Visual inspection (VI) is the most basic type of NDT technique to inspect damage. It is quick, economically viable, and flexible, while its disadvantages are quite obvious and significant [11]. VI methods include visual & optical testing (VOT) and penetrant testing (PT). VOT analysis is a leading procedure in the monitoring of surface imperfections for acceptance-rejection criteria during composite parts production [52]. The PT technique is a widely applied, low-cost inspection method. It has been used in non-porous materials to detect casting, forging, and welding surface defects including cracks, surface porosity, leaks in new products, and in-service fatigue cracks *etc*.

VI methods are particularly effective in detecting macroscopic flaws, such as poor joints, erroneous dimensions, poor surface finish and poor fits. It usually employs easy-to-handle equipment such as miniature cameras or endoscopes [23]. VI studies of small integrated circuits have shown that the modal duration of eye fixation from trained inspectors was ~200 ms. Here, variation by a factor of six in inspection speed led to a variation of less than a factor of two in inspection accuracy; inspection accuracy also depends on training, inspection procedures, and apparatus (optics, lighting, *etc*.) [53].

## 3.2    Acoustic emission

Damage occurrences within a composite produce localised transient changes in stored elastic energy; the energy releases stress waves, resulting in fibre breakage, matrix cracking, debonding, delamination *etc*. [3]. Acoustic emission (AE) based NDT techniques detect and track these sudden releases of stress waves through arrays of highly sensitive sensors or transducers [54], as illustrated in Figure 5. Use of the AE method started in the early 1950s, when Kaiser first used electronic instrumentation to detect audible sounds produced by tensile deformation of a metallic specimen [55]. His discovery on the effect of sample stress history on the production of AE became known as the 'Kaiser effect' [56]. AE was first applied to the study of composite materials in the 1970s [3], and it has now been widely used in various aspects of composite research [57].

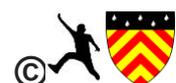

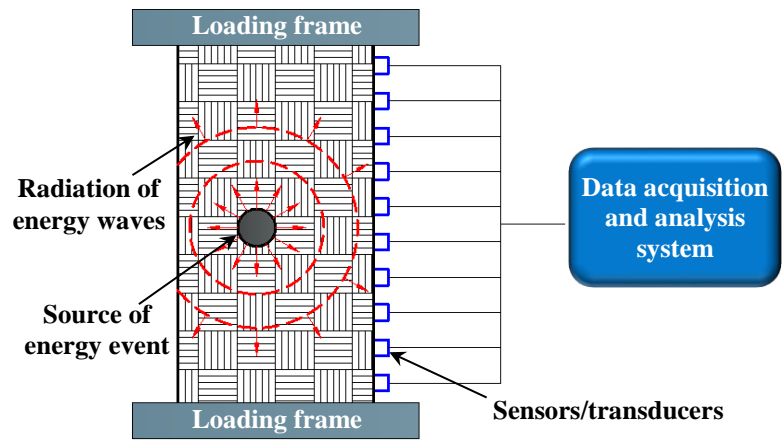

**Figure 5**     **Schematic of localised transient changes in stored elastic energy within a material system under loading, showing the measuring principle of the acoustic emission based NDT technique.**

The AE method is unique in that: (i) the signals, i.e. stress waves, are emitted by the testing sample, not from external sources (as with other NDT methods); (ii) strain or displacement data are usually recorded, rather than as geometrical defects; (iii) it monitors dynamic processes in a material, tracking the development of certain defects, which significantly benefits fatigue tests. It has been reported that AE-based NDT can detect fatigue cracks, fibre fractures, matrix micro-cracks, interface debonding, as well as delamination [11]. However, there are also certain difficulties. Data collected during the loading of a composite system can be in different forms, the most common is the AE amplitude signal. Processing and analysis of data are time-consuming and require certain skills and experience [3]: in particular, the distribution of amplitudes exhibit overlapped areas, which sometimes causes difficulties in associating these with the damage mechanisms.

Efforts have been made to address these issues. A common approach is to analyse multiple parameters to complement the damage analysis, such as cumulated event counts [58,59], energy [60], duration [61], or frequency of the received amplitude signals [62]. Other solutions include verifying damage modes through other methods, for example, microscopy, to provide more reliable analysis [63]. Table 1 summarises some of the commercial suppliers of AE-based NDT systems, which may be applied to composite research. Standardised practices of using AE include: ASTM E1067 on examining GFRP tanks/vessels [64]; ASTM E1118 on composite pipes [65]; ASTM E2191 on filament-wound composite pressure vessels [66]; ASTM E2076 on composite fan blades [67]; as well as ASTM E2661 on plate-like and flat composite structures for aerospace [68].

**Table 1**     **Summary of suppliers of devices and systems used for the acoustic emission-based NDT technique; parameters are extracted from literature.**

| Supplier | Resolution (bit) | Dynamic range (dB) | Bandwidth (MHz) | Sampling speed (MHz) | Ref |
|---|---|---|---|---|---|
| Physical Acoustics Co. | 16 | 100 | 0.001-1.2 | 5-20 | [69–71] |
| Meggitt Endevco | -- | 100 | 0.002-1.0 | -- | [63] |
| Soundwel Technology Co. Ltd | 16 | 85 | 0.001-2.5 | 0.5-10 | [72] |
| IPPT PAN | -- | 40 | 0.005-0.5 | -- | [73] |
| Vallen Systeme | -- | 100 | 0.1-0.45 | -- | [74] |

*Please note the information in this table is incomplete, and not for advertising purposes – it should not be taken as endorsements by the authors.

There is also some interest in a combined method of acoustic emission and ultrasonic testing, namely the acousto-ultrasonic technique (AUT), as first introduced by Vary in 1981 [75]. By adopting the ultrasonic transducer, repeated ultrasonic pulses are introduced into a material, resultant waveforms carry the morphological information that contribute to damage mechanisms. A concept of 'stress wave factor' is defined as a relative measurement of efficiency of energy dissipation to indicate regions of damage [76]. In NDT, the AUT is mainly used to determine the severity of internal imperfections and inhomogeneities in composites [11].

## 3.3     Ultrasonic testing

Ultrasonic testing (UT) is an acoustic inspection technique, which is expanding rapidly into many areas of manufacturing and in-service detection [77]. It operates through surface wave testing, bulk wave propagation and guided wave propagation, while the guided wave analysis technique is superior for anisotropic materials [78]. For further information, the use of ultrasonic bulk wave testing in sizing of flaws has been reviewed by Felice and Fan [79].

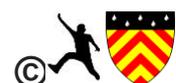

For NDT inspection of composite materials, elastic waves or 'Lamb waves' propagate in selective directions owing to their anisotropic nature which makes the technique effective. UT operates in three detection modes, i.e. reflection, transmission and back-scattering of pulsed elastic waves in a material system [17]. It introduces guided high frequency sound waves (ranging from 1 KHz to 30 MHz [4]) to effectively detect flaw size, crack location, delamination location [80], fibre waviness [31], meso-scale ply fibre orientation [81], and layup stacking sequence [82].

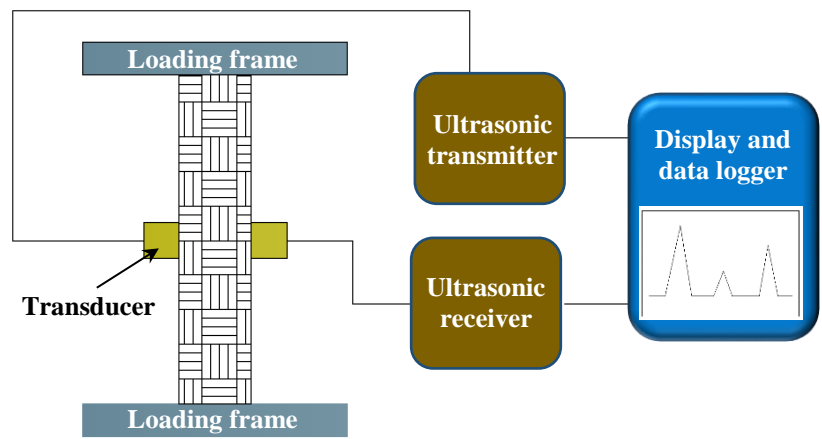

**Figure 6**      **Principle of ultrasonic testing a composite material in transmission mode.**

There are various types of UT systems with hundreds of guided wave modes and frequencies being available [78]. A typical UT system consists of a transmitter and receiver circuit, transducer tool, and display devices, see Figure 6. The transmitter can either be arranged at an angle to the sample, or in the form of phased array [83]. The guided Lamb waves can be generated by means of (i) ultrasonic probe, (ii) laser, (iii) piezoelectric element, (iv) interdigital transducer, or (v) optical fibre [84].

The potential types of damage that guided Lamb wave-based NDT can provide are summarised by Rose [77]; the mode selection, generation and collection, modelling and simulation, signal processing and interpretation have been well-documented by Su *et al*. [84]. A later review on guided waves for damage identification in pipeline structures is provided by Guan [85]. Table 2 provides some suppliers of UT equipment which may be applied to composites research. UT techniques for composites have been standardised: ASTM E2373 gives the requirements for developing a time-of-flight UT examination [86]; ASTM E2580 for inspections on flat composite panels and sandwich structures in aerospace applications [87]; ASTM E2981 for filament wound pressure vessels in aerospace applications [88].

**Table 2**      **Summary of suppliers for the ultrasonic-based NDT technique; parameters are extracted from published references as practical guidance to their applications to composite research.**

| Supplier | Resolution (bit) | Dynamic range (dB) | Bandwidth (MHz) | Sampling speed (MHz) | Ref |
|---|---|---|---|---|---|
| ZETEC Inc. | 16 | -- | 0.5-18 | 50 or 60 | [89] |
| Inspection Technology Europe | 16 | 90 | 0.1-30 | 50/160 | [90] |
| Advanced Technology Group | 12 | 80 | 1-22 | 100 | [91] |
| Peak NDT | 16 | 60 | 0.001-40 | 10-100 | [92] |
| Olympus | -- | 60 | 0.2-20 | -- | [93] |
| Polytec Co. | 14 | | 0-25 | | [94] |

*Please note the information in this table is incomplete, and not for advertising purposes – it should not be taken as endorsements by the authors.

## 3.4      Infrared thermography

Infrared thermography (IRT) is a method used to detect and process infrared energy emissions from an object by measuring and mapping thermal distributions [95]. Infrared energy is electromagnetic radiation with wavelengths longer than visible light, see Figure 3. The discovery of thermal radiation dated back to the early 1800s [52]. Every object with a temperature higher than absolute zero emits electromagnetic radiation that falls into the infrared spectrum [96]. IRT has undergone rapid development in the last 30 years with developments in infrared cameras, data acquisition and processing techniques. It provides capabilities in terms of non-contact, non-invasive, real-time measurement, high resolution, and covering large volumes [97].

The IRT method is effectively used to monitor the entire life of a product, from manufacturing (on-line process control), to the finished product (NDT evaluation) and to in-service maintenance and diagnostics [52]. It has been applied to research and various aspects within industry, including non-destructive testing [98], building diagnostics [99], adhesion science [52], condition monitoring [100], predictive or preventive maintenance [101], medical diagnostics [102], veterinary medicine [103], and many more. As for composite materials and structures, IRT-based NDT has also been widely used, especially during manufacturing for aerospace applications. It is used to detect inclusions, debonding, delamination, and cracked networks [27]. Both Boeing and Airbus have used IRT for structural health monitoring to ensure the integrity of their composite products [97].

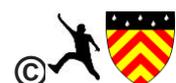

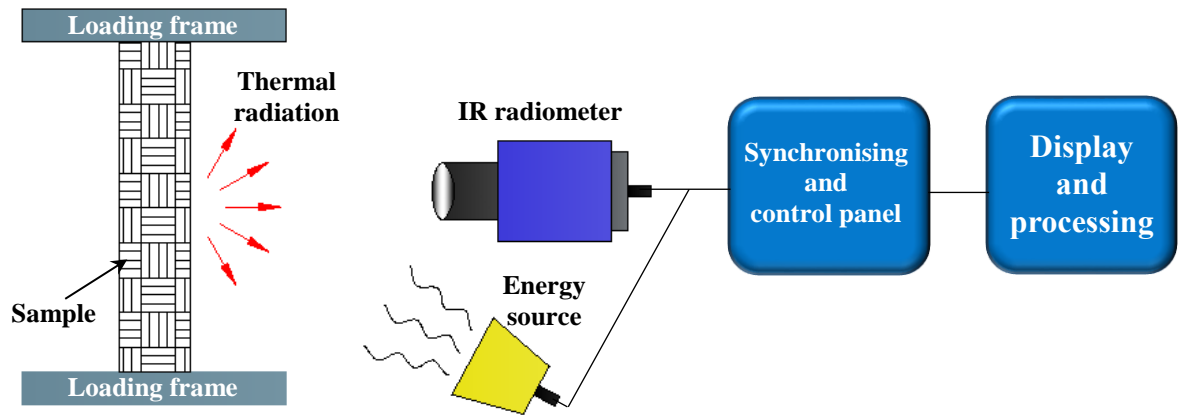

**Figure 7**     **Schematic of the measurement principles for an infrared thermography system in reflection mode.**

A typical IRT system contains an infrared radiometer, with/without energy source, synchronising and control panel, display software, see Figure 7. The radiometer is the core of the IRT system, it absorbs IR energy emissions and converts them into electrical voltage or current signals. They are then transmitted and displayed as infrared images of temperature distribution [52]. The use of IRT can be implemented through (i) passive and (ii) active thermography [104]. In passive thermography (PT), thermal radiation is directly emitted from surfaces of the test body under natural conditions and subsequently monitored. For active thermography (AT), a heating or cooling flow is generated and propagated into the test object, thermal responses according to the Stefan-Boltzmann law are then detected and recorded to reveal internal structures. Recent advances in signal processing techniques and equipment developments have made the AT method more practical and effective than the conventional PT approach [105,106].

Based on energy stimulation methods, the AT method has developed into different categories. First, optical thermography is the most traditional form of IRT, using optical sources such as photographic flashes, halogen lamps, or lasers, which are also known as pulsed thermography (PT) [107], modulated (lock-in) thermography (MT) [108], or laser thermography (LT) [109,110], respectively. Second, induction thermography, which shares similar principles to eddy current testing, that uses electronic or magnetic currents to induce energy waves [111–114]. Third, mechanical thermography, which uses mechanical waves to interact with internal structures to detect thermal waves from defects [115]; it can be implemented through vibrothermography [116,117], microwave thermography [118,119], or ultrasonic lock-in thermography (ULT) [120] which attracts increasing interest. Yang and He [121] have presented a comprehensive review of the optical and non-optical IRT methods and their NDT applications in composite materials/structures. Table 3 gives some suppliers of IRT equipment which may be applied to composites research. The reader is referred to ASTM E2582 for standard practice on using IRT with composite panels and repair patches in aerospace applications [122].

**Table 3**    **Summary of commercial infrared thermography systems and their key parameters applied to composite research.**

| Supplier | Spatial resolution (mRad) | Thermal sensitivity (mK) | Imaging resolution (pixel²) | Imaging rate (fps) | Ref |
|---|---|---|---|---|---|
| Thermal Wave Imaging Inc. | 1.13 | 25 | 320×256 | 60 | [123,124] |
| Thermoteknix System Ltd. | 0.47 | 70 | 384×288 | 50 or 60 | [125] |
| Fluke Corporation | 0.93 | 50 | 640×480 | 9 or 60 | [126] |
| InfraTec GmbH | 0.08-1.3 | 20 | 640×512 | 1-100 | [127] |
| Mikron Infrared | 1.0 | 80 | 320×240 | 9 or 50 | [128] |
| Optris | -- | 130 | 382×288 | 80 | [129] |
| NEC Avio | 0.87 | 50 | 320×240 | 60 | [130] |
| FLIR System | 0.19-1.36 | 20 | 320×256 | 50 | [131,132] |

\*Please note the information in this table is incomplete, and not for advertising purposes – it should not be taken as endorsements by the authors.

## 3.5    Terahertz testing

Terahertz (THz) waves lie within the electromagnetic spectrum from 100 GHz to 30 THz [133], which belong to non-ionising radiation and are not harmful to biological tissues (Figure 3). There are many THz wave sources in nature, though previously it has been difficult to generate and detect THz waves, so for many years, there have been few applications [134]. Owing to breakthroughs in ultrafast lasers and ultra-micro machining technologies during the 1980s [135,136], there has been a rapid expansion in applications for THz science and technology [46]. THz-based NDT technology has also started to be a promising technique for composite inspections [2,47], offering advantages in terms of higher resolution and better penetration in most materials compared to other techniques [137].

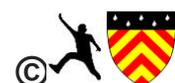

THz waves have good penetrating power for non-metallic, non-polar materials, including foams, ceramics, glass, resin, paint, rubber and composites. THz-based NDT techniques use the wave characteristics to detect, analyse and evaluate material systems, which has attracted wide interest in various fields, leading to rapid expansion [138]. Figure 8 shows an example of a typical setup of the THz-based NDT method, presenting the basic measuring principles in transmission mode [45]. The system induces THz short waves into a material, which interact with different phases, inclusions, defects or damage. Internal structures within the material are determined by detection and analysis of reflected or transmitted THz waves. Therefore, the multi-phase and multi-layered nature of composites are well-suited to THz-based NDT – it offers multi-scale, more comprehensive information to detect and reveal internal structures and damage within a composite [47].

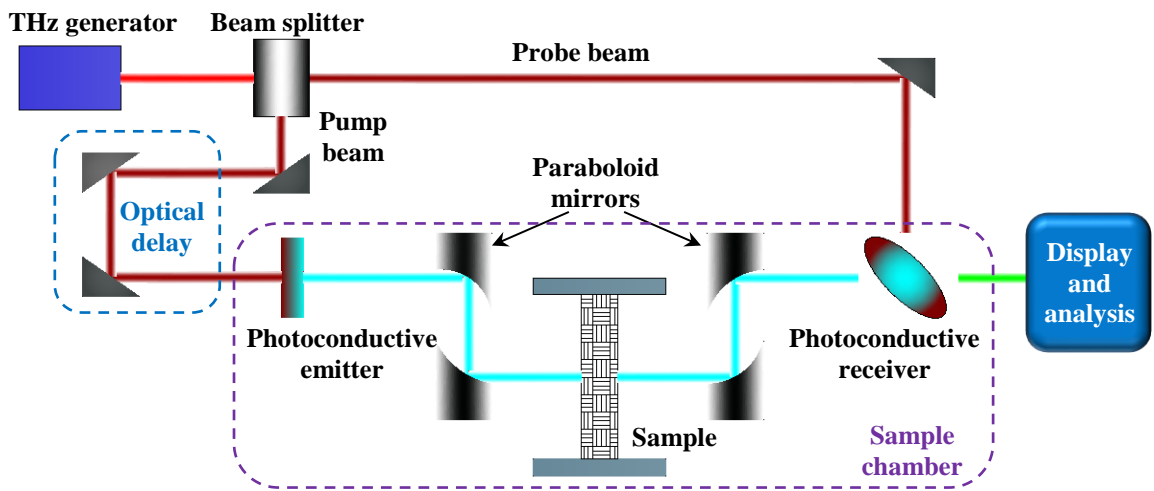

**Figure 8**       **Schematic showing the measurement principles of THz time-domain spectroscopy using photoconductive antennas.**

The THz-based NDT technique is usually implemented through (i) a THz time-domain spectroscopy system (THz-TDS), also known as pulsed spectroscopy, or (ii) a continuous wave (THz-CW) system. The detection setup determines how information is evaluated within composite materials. In the THz-TDS system, short pulsed THz waves are generated by optical excitation of a photo-conductive antenna using a laser pulse emitting in the femtosecond regime [133], the time-dependent evolution of the THz electric field of a single pulse is measured, which can be used to determine phase information within a composite. For the THZ-CW system, high power THz waves are produced through gas lasers, quantum cascade lasers or parametric sources [2], and phase information is measured by recording the average intensity (related to the amplitude of the wave) of the electromagnetic field.

**Table 4**       **Summary of THz-based NDT system suppliers; key parameters are adopted as examples from published research.**

| THz supplier | Setup | Resolution (bit) | Dynamic range (dB) | Scanning range (mm²) | Spectrum band (THz) | Ref |
|---|---|---|---|---|---|---|
| Virginia Diodes Inc. | THz-CW | -- | 110 | 100×100 | 0.05-1.0 | [137] |
| Zomega Terahertz Co. | THz-TDS | -- | 70 | 150×150 | 0.1-4.0 | [139,140] |
| TeraView | THz-TDS | 16 | 95 | 150×150 | 0.06-3.0 | [141] |
| TeraSense | THz-CW | 16 | -- | 128×64 | 0.05-1.0 | [142] |
| MenloSystems | THz-TDS | 18 | 90 | 150×150 | 0.1-4.0 | [143] |
| Toptica Photonics | THz-CW | -- | 100 | 100×100 | 0.1-6.0 | [144] |
| Luna Ltd. | THz-TDS | 16 | 95 | -- | 0.1-2.0 | [145–147] |

*Please note the information in this table is incomplete, and not for advertising purposes – it should not be taken as endorsements by the authors.

There have been several reviews regarding applications of the THz-based technique, focusing on different aspects: Dhillon *et al*. [46] presented a comprehensive review on the roadmap of THz science and technology; Jansen *et al*. [148] reviewed progress and applications of THz systems in the polymer industry; Amenabar *et al*. [2] summarised the detection and imaging methods using THz waves, as well as their applications in composites; Zhong [47] further summed up the most recent advances. Table 4 shows some of the commercialised THz-based NDT systems and their key parameters. As an emerging NDT technique, standardised practice on using the THz approach is still developing.

## 3.6     Shearography

Shearography testing (ST) is a laser-based non-contact NDT technique, using a full-field speckle shearing interferometric method to overcome the limitations of holography testing [49]. This technique was first described and applied by Leendertz in the 1970s [149,150]. To date, it has been used in various fields as a practical quantitative inspection tool to detect flaws and defects [151–153], leakage [154], delamination and damage [155,156], as well as measurement of displacement and strain [157,158], curvature [159–162], residual stress [163–

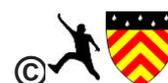

165], mechanical analysis [166,167], surface profiling [168], and dynamic vibration [169–171].

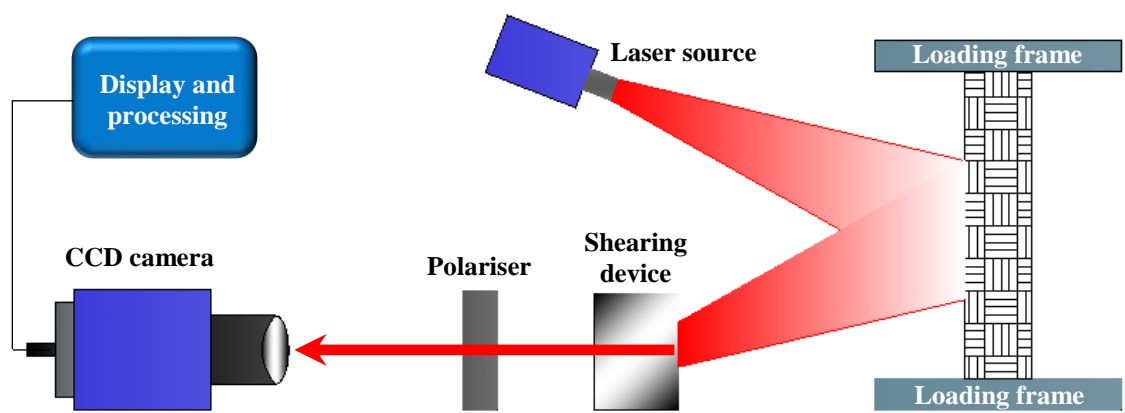

**Figure 9**    Schematic illustration of a shearography system.

**Table 5**    Summary of commercialised shearography system suppliers, and their key parameters applied to composite research.

| Supplier | Hardware & software | Inspection area (m²) | Imaging resolution (pixel²) | Imaging rate (Hz) | Ref |
|---|---|---|---|---|---|
| Dantec Dynamics | Yes | 0.01 – 2 | 1392×1040 | 10 | [91,172] |
| ZEISS Optotechnik | Yes | -- | 220×160 | -- | [173] |
| Optonor AS | Yes | 0.01 – 4 | 1936×1216 | -- | [174,175] |
| ISI-sys | Yes | -- | 2560×1920 | -- | [176,177] |
| Laser Optical Engineering Ltd. | Yes | 0.49 | 1280×1024 | 12 | [178] |
| Laser Technology Inc. | Yes | 0.01-0.05 | 1628×1236 | 30 | [179] |

\*Please note the information in this table is incomplete, and not for advertising purposes – it should not be taken as endorsements by the authors.

A typical shearography setup is shown in Figure 9. A laser beam illuminates a sample surface, and the beam is then scattered and reflected. The resulting speckle pattern is imaged through a shearing device (Michelson interferometer or diffractive optical element), which divides it into two coherent images with one being monitored during deformation. A controlled stressing process is essential and is applied through thermal [180,181], vacuum [163,182], vibration [153], microwave [183], or mechanical loading [184]. The interferometric pattern is then captured and recorded by a charge-coupled device (CCD camera) which results in a fringe pattern that contains structural information [178]. It has been adopted for inspection and evaluation in various composite products, for example pipes [185,186], sandwich structures [16,17,187], wind turbine blades [188], aerospace structures [189–191], as well as racing tyres [192]. Some suppliers of commercialised shearography systems are given in Table 5. An example of standard practice using shearography for polymer composites and sandwich core materials in aerospace is represented by ASTM E2581 [187].

## 3.7      Digital image correlation

Digital image correlation (DIC) is a simple and cost effective optical NDT technique for measuring strain and displacement, which are critical parameters within engineering and construction projects. It was developed in the 1980s [193], and has become widely used only in recent years due to the rapid development of computers and image acquisition methods. Images are usually captured through CCD cameras, possibly with the aid of microscopy. The DIC system tracks blocks of random pixels on a sample surface, and compares digital photographs at different stages of deformation to build up full field 2D or 3D deformation vector fields and strain maps [194]. Thus, any changes in the structure or surface can easily be reflected to give details on surface strain, deformation, or crack propagation, making it ideal for studies of crack propagation and deformation. It offers more accurate strain monitoring than conventional extensometers or strain-gauges, which often suffer from imperfect attachment to the measured surface and the limitations imposed by values that are averaged over the gauge length [195].

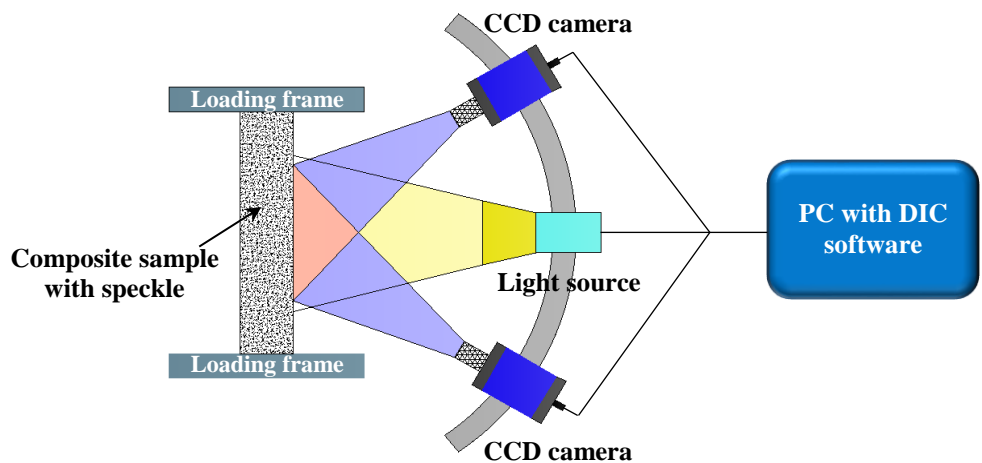

**Figure 10      Schematic of a typical stereo-DIC setup for strain mapping of a composite sample sprayed with stochastic speckle pattern.**

Figure 10 shows a typical DIC system for strain mapping of a composite sample; here, special illumination may be required. The sample is sprayed with a white stochastic speckle pattern prior to testing and two CCD cameras need to be calibrated each time. Imaging data can be analysed through commercialised software to reveal changes in speckle with reference images and strain or deformation can be calculated during the tests. Thus, quality of the speckle pattern is vital for accuracy and precision in the DIC technique [196]. Pan [197] presented the historic developments, recent advances and future of DIC for surface deformation measurement; Hild *et al* [198] discussed the capabilities of DIC in damage measurements; Aparna *et al* [199] summarised studies on fatigue testing of composites using the DIC technique. Therefore, they are not elaborated here.

Table 6 summarises some suppliers of DIC systems and certain examples in the literature. Audiences are recommended to refer to each supplier for full details. Given the flexibility and versatility of DIC systems, standardisation of the DIC technique is difficult or even impossible to be applicable to each individual situation [197].

**Table 6**     **Summary of suppliers of DIC systems and their key parameters applied to composite research.**

| DIC supplier | Hardware/ Software | Maximum imaging rate (fps) | Imaging resolution (pixel²) | Precision (μm/pixel) | Strain mapping | Ref |
|---|---|---|---|---|---|---|
| Correlated Solutions | Yes/Yes | 5,000,000 | 2240×1680 | -- | 2D & 3D | [200] |
| Limess | Yes/Yes | 2,000,000 | 4096×3068 | 1.0 | 2D & 3D | [201] |
| Dantec Dynamics | Yes/Yes | 1,000,000 | 2560×1920 | 5.0 | 2D & 3D | [202] |
| GOM | Yes/Yes | 2000 | 4096×3068 | 4.8 | 2D & 3D | [203–205] |
| LaVision | Yes/Yes | 150 | 2240×1680 | 5.86 | 2D & 3D | [206] |
| Instron | Yes/Yes | 50 | 1280×720 | 1.0 | 2D | [207] |
| MatchID | No/Yes | -- | -- | -- | 2D & 3D | [208] |
| ISI-sys GmbH | Yes/No | 100 | 1920*1200 | 2.0 | 2D & 3D | [209] |

*Please note the information in this table is incomplete, and not for advertising purposes – it should not be taken as endorsements by the authors.

## 3.8     Imaging techniques

Imaging techniques refer to the NDT methods that are based on phase contrast imaging, which were first developed in the 1930s [210]. They enable high-resolution imaging (a few angstroms), making it possible to distinguish features at atomic or molecular levels. Developments in digital imaging technology and synchrotron radiation facilities have promoted the use of imaging techniques since the 1990s [211]. To date, it has been reported that X-ray imaging is carried out either through lab-based X-ray tubes or synchrotron light sources; alternatively, neutron imaging uses neutrons generated from fission reactors or spallation sources [212]. Both X-ray and neutron radiography have developed to be indispensable tools in various research fields ranging from solid matter to soft tissues [211].

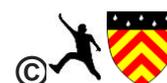

Synchrotron X-ray and neutron radiation are NDT techniques that provide insights into micro-structures, residual stress, strain and stress fields, crystallographic textures *etc*., at atomic and crystalline levels. Their measurement and detection principles are similar, mainly depending on scattering techniques, see Figure 11. The incident light beams (monochromatic or white) are directed onto a sample, the scattered beams are captured by the detectors as a function of momentum transfer and/or transferred energy $\Delta E$ [213]. Diffraction patterns from a material in (a), can be used to characterise the crystalline structure, residual stress, and crystallographic textures; in (b), small-angle scattering (SAS) uses smaller scattering angles than (a) to investigate material structures with various substances to provide quantitative statistical information at nanoscale levels; in (c), reflectometry is used to study the surface morphology of thin films or multi-layered composites *etc*. [214]; in (d), spectroscopy is performed to determine electronic, vibrational or magnetic excitations, and diffusional processes in solids and liquids.

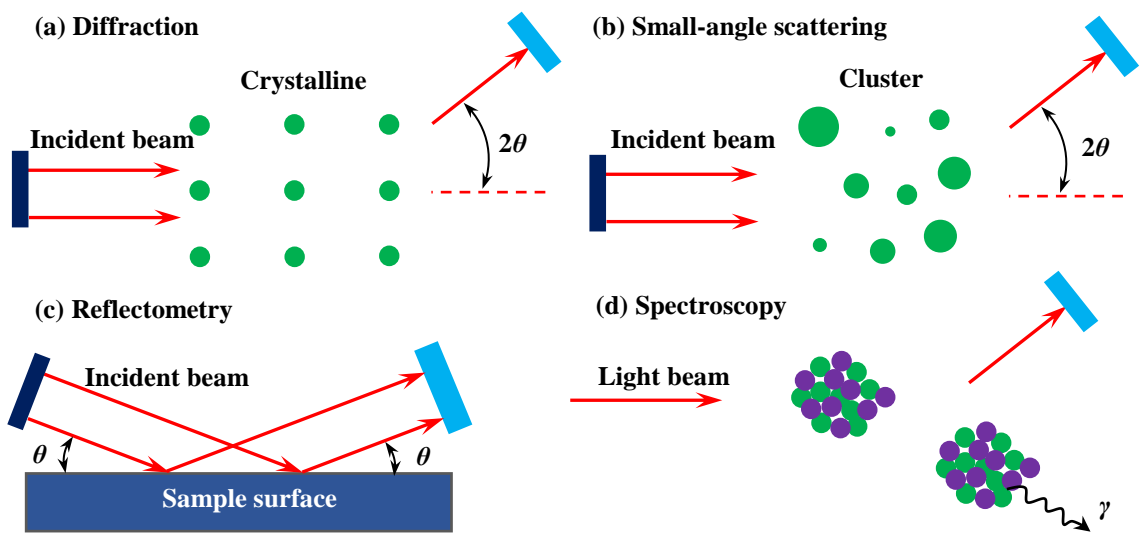

**Figure 11**    **Schematic of synchrotron and neutron measurement techniques: (a) diffraction; (b) small-angle scattering; (c) reflectometry; (d) spectroscopy.**

Although synchrotron and neutron imaging share basic principles, the neutron technique is superior in terms of penetration depth; i.e. X-rays (photons) can only be used non-destructively for examination in the near-surface regions [215]. Neutrons carry no electric charge, so there is no electrostatic interaction with the electron cloud of an atom [213]. The characterisation and analysis of residual stresses in materials science, by using synchrotron and neutron radiation has been documented by Fitpatrick and Lodini [213], and Hutchings *et al*. [216]. Also Banhard *et al*. [212] have reviewed the applications of X-ray and neutron imaging to materials science and engineering.

### 3.8.1    X-ray imaging

A commonly used laboratory-based X-ray source for imaging is the X-ray tube, as schematically illustrated in Figure 12. A bias voltage of 30-60 kV is applied between the filament and metallic target in an evacuated X-ray tube, causing electrons emitted from the filament to collide with the metallic target at high velocity (energy) and radiate X-rays. The wavelengths of the X-rays are about 0.5-2.5 Å, and depend on the target material. The most commonly used metallic target is copper, which emits strong X-rays with a wavelength of 1.54 Å [217]. Laboratory X-ray imaging (XRI) systems are usually cheaper and easier to access, and are suitable for materials with higher phase contrast, such as glass fibre reinforced composites. The acquisition time for laboratory XRI ranges from minutes at low resolution (sub-millimetre) to hours or even days at high resolution (sub-micron) [218].

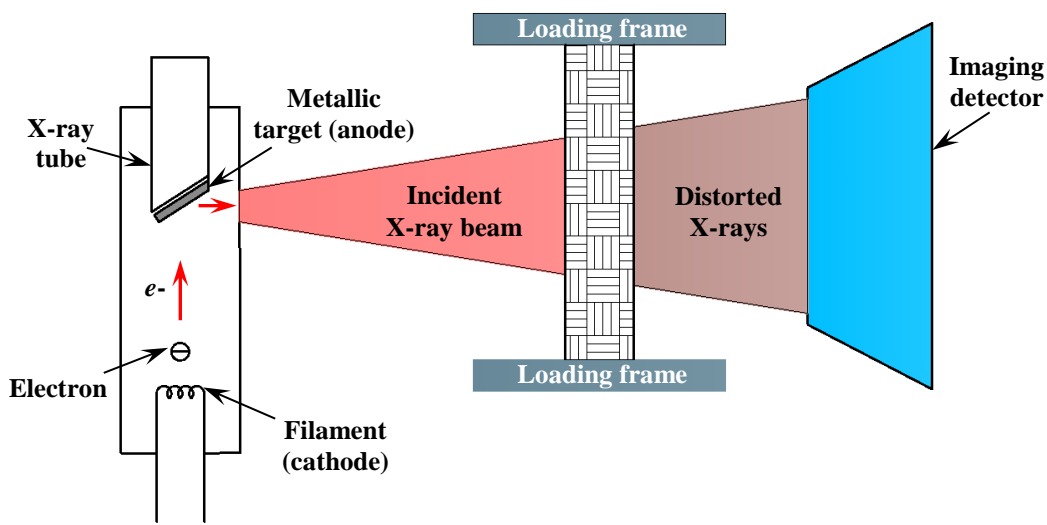

**Figure 12        Schematic representation of laboratory-based X-ray imaging.**

A major disadvantage of laboratory XRI systems is the lack of capability to penetrate deeply into engineering materials, which depends on X-ray energy and wavelength [216].

Although gamma-rays have higher penetrating capacity than X-rays, they are usually generated from a radioactive source, which cannot be turned off and is difficult to adopt as a compact source to provide a photon flux comparable to an X-ray tube; thus detection efficiency is fairly low and long measuring times are required [219].

The limitations on penetration depth have been overcome by the rapid development of synchrotron facilities. Laboratory X-ray sources produce polychromatic and divergent X-ray beams, while a synchrotron X-ray beam is parallel, monochromatic, more coherent, with higher orders of flux and brightness. These factors determine the image quality and acquisition time. A synchrotron XRI system offers much higher levels of both signal-to-noise ratio and phase contrast, which makes it superior for low contrast materials, such as carbon fibre reinforced composites. The high flux and brilliance of the X-ray beam allows very fast imaging acquisition with high resolution; *e.g.* 1 tomogram per second with 1.1 μm spatial resolution using the TOMCAT beamline at the Swiss Light Source facility [220].

Synchrotron facilities have gone through four generations of technical evolution. The first-generation synchrotron facility was built in the US in 1946 and was primarily used for high-energy particle physics. Second-generation synchrotrons were dedicated to the production of synchrotron light in the 1980s, which used bending magnets to generate synchrotron light. Third-generation light sources originated in the 1990s [221], with facilities that used insertion devices (wigglers and undulators) to produce intense and tuneable X-ray beams. Fourth-generation facilities will be based on free electron lasers which offer more advanced capabilities to generate brighter light sources [222]. Currently, there are about 50 synchrotron facilities around the world, supporting various investigations in engineering, health and medicine, materials science, chemistry, cultural heritage, environmental science and many more [222–231]. Table 7 summarises the 3rd- and 4th-generation synchrotron light source facilities throughout the world. Given that the first 3rd-generation synchrotron facilities were built in 1993, some will be subjected to upgrading in the near future [231].

XRI can also be implemented through different methods as recently presented by Liu *et al.* [210]; Garcea *et al.* reviewed the applications of X-ray computed tomography (CT) to polymer composites [218]. Standard practice in using computed radiography (X-rays or γ-rays) for metallic and non-metallic materials is recommended in ASTM E2033 [232]; ASTM

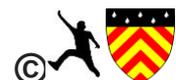

E2662 provides guidance on the radiographic examination of flat composite panels and sandwich core materials for aerospace applications [233].

**Table 7**      **Summary of third and fourth-generation synchrotron light sources in operation and under construction world-wide; brilliance (brightness) is shown in photons/s/mrad$^2$/mm$^2$/0.1%bw.**

| Country | Location | Source | Energy (GeV) | Brilliance | Circumference (m) | Beamlines | Operation year/status |
|---|---|---|---|---|---|---|---|
| US | Berkeley | ALS | 1.9 | $10^{19}$ | 196.8 | 40 | 1993 |
| Italy | Basovizza | ELETTRA | 2.0/2.4 | $10^{19}$ | 259.2 | 28 | 1994 |
| France | Grenoble | ESRF | 6.0 | $10^{19}$ | 844 | 56 | 1994 |
| China | Taiwan | TLS | 1.5 | $10^{17}$ | 120 | 25 | 1994 |
| Korea | Pohang | PLS | 3.0 | $10^{20}$ | 281.82 | 36 | 1995 |
| US | Lemont | APS | 7.0 | $10^{14}$ | 1104 | 68 | 1995 |
| Japan | Hyogo | SPing8 | 8.0 | $10^{20}$ | 1436 | 62 | 1997 |
| Germany | Berlin | BESSY II | 1.7 | $10^{19}$ | 240 | 46 | 1999 |
| Canada | Saskatoon | CLS | 2.9 | $10^{20}$ | 171 | 22 | 1999 |
| Switzerland | Villigen | SLS | 2.4 | $10^{20}$ | 288 | 16 | 2001 |
| US | Menlo Park | SPEAR3 | 3.0 | $10^{20}$ | 234 | 30 | 2004 |
| UK | Didcot | DLS | 3.0 | $10^{20}$ | 561.6 | 39 | 2006 |
| France | Saint-Aubin | SOLEIL | 2.75 | $10^{19}$ | 354 | 29 | 2006 |
| Australia | Clayton | AS | 3.0 | $10^{12}$ | 216 | 10 | 2007 |
| China | Beijing | BEPC II | 2.0 | $10^{13}$ | 240 | 14 | 2008 |
| Germany | Hamburg | PETRA III | 6.0 | $10^{21}$ | 2304 | 21 | 2009 |
| China | Shanghai | SSRF | 3.5 | $10^{20}$ | 432 | 31 | 2009 |
| Span | Barcelona | ALBA | 3.0 | $10^{20}$ | 268.8 | 9 | 2012 |
| US | Upton | NSLS-II | 3.0 | $10^{21}$ | 792 | 28 | 2014 |
| China | Taiwan | TPS | 3.0 | $10^{21}$ | 518.4 | 7 | 2016 |
| Krakow | Poland | SOLARIS | 1.5 | $10^{18}$ | 96 | 2 | 2016 |
| Jordan | Allan | SESAME | 2.5 | $10^{18}$ | 133 | 7 | 2017 |
| Germany | Schenefeld | XFEL | 17.5 | $10^{33}$ | 1700 | 6 | 2017 |
| US | Ithaca | CHESS-U | 6.5 | $10^{21}$ | 768.4 | 11 | 2018 |
| Sweden | Lund | MAX IV | 3.0 | $10^{22}$ | 528 | 17 | 2019 |
| France | Grenoble | ESRF-EBS | 6.0 | $10^{21}$ | 844 | 8 | Under construction |
| Brazil | Campinas | SIRIUS | 3.0 | $10^{21}$ | 518.4 | 13 | Under construction |
| China | Beijing | HEPS | 6.0 | $10^{22}$ | 1360 | 14 | Under construction |

### 3.8.2    Neutron imaging

The neutron was discovered by Sir James Chadwick at Cambridge in 1932 [234] through the collision of beryllium by $\alpha$-particles from polonium. Neutrons have a wave character, their wave-lengths are in the order of interatomic distances (~0.1 nm), and kinetic energies close to atomic vibration energies (~$10^{-2}$ eV). Thus, they give rise to the possibilities of diffraction and inelastic scattering studies, which were experimentally demonstrated in 1946 by Wollan and Clifford using the Graphite Reactor at Oak Ridge National Laboratory, in the era of the Manhattan project in the US [213]. Important progress was made on neutron strain scanning (NSS) during the 1960s and 1970s. Techniques such as small-angle neutron scattering (SANS), neutron time-of-flight (TOF) scattering, backscattering or spin-echo techniques and neutron reflectivity subsequently broadened the applications of NSS to larger scientific domains such as solid state chemistry, liquids, soft matter, materials science, geosciences and biology [235].

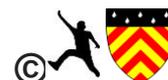

A schematic example of NSS using ENGIN-X (ISIS Neutron and Muon Source, Rutherford Appleton Laboratory, UK) is presented in Figure 13 [236]. The pulsed neutron beam with a wide range of energy travels to the sample and, being scattered, the detectors then collect the diffracted neutrons at a fixed angle of $2\theta_b$. As neutrons can penetrate deeply into composite materials, strain/stress can be non-destructively measured [237]. Neutrons are difficult and expensive to produce – neutron sources are usually generated through either nuclear fission reactors (continuous neutron sources) or neutron spallation (pulsed neutron sources). Neutron source facilities are summarised chronologically in Table 8 [213,238,239].

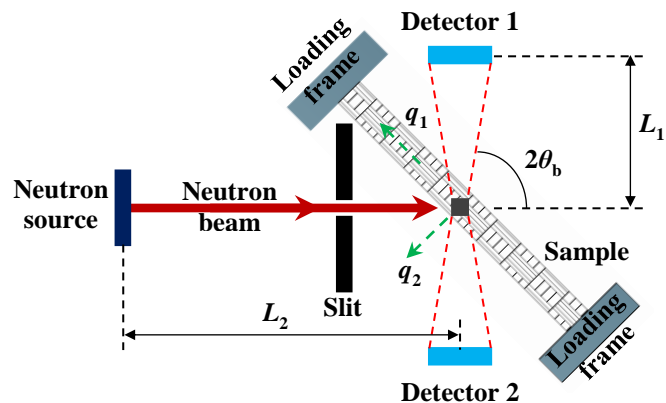

**Figure 13**     **Schematic representation of a time-of-flight neutron strain scanner at the ENGIN-X. The elastic strain is measured along the directions of the impulse exchange vectors, $q_1$ and $q_2$, through the two detectors.**

Neutron imaging (NI) has progressed as a reliable NDT technique, in the forms of neutron topography and radiography; specialised instrumentation at pulsed neutron sources include RADEN@J-PARC [240] and IMAT@ISIS [241]. Neutron tomography allows 2D or 3D imaging of the attenuation coefficient distribution within a material system, thus internal structures and material composition can be visualised [242]; neutron radiography is a transmission imaging technique for heterogeneous materials, taking advantage of the scattering and/or absorption contrast between different elements [243].

**Table 8**     **Summary of neutron source facilities world-wide and their basic parameters.**

| Country | Location | Source | First open | Power (MW) | Flux ($10^{14}$ n/cm$^2$/s) | Scattering instruments | Operating time (d/year) |
|---|---|---|---|---|---|---|---|
| *Continuous neutron sources* | | | | | | | |
| Canada | Chalk River | NRU | 1957 | 120 | 3.0 | 6 | 300 |
| Australia | Lucas Heights | HIFAR | 1958 | 10 | 1.4 | 7 | 300 |
| Hungary | Budapest | BNC | 1959 | 10 | 1.6 | 7 | 200 |
| Russia | Gatchina | WWR-M | 1959 | 16 | 1.0 | 12 | 200 |
| Denmark | Risø | DR3 | 1960 | 10 | 1.5 | 7 | 286 |
| Sweden | Studsvik | R-2 | 1960 | 50 | 4.0 | 8 | 187 |
| Japan | Tokai | JRR-3 | 1962 | 20 | 2.0 | 23 | 182 |
| Germany | Juelich | FRJ-2 | 1962 | 23 | 2.0 | 16 | 200 |
| Netherlands | Delft | HOR | 1963 | 2 | 0.2 | 11 | 160 |
| US | Brookhaven | HFBR | 1965 | 30 | 4.0 | 14 | 260 |
| US | Columbia | MURR | 1966 | 10 | 6 | 4 | 338 |
| Russia | Ekaterinburg | IWW-2M | 1966 | 15 | 2.0 | 6 | 250 |
| US | Oak Ridge | HFIR | 1966 | 85 | 25.0 | 14 | 210 |
| Norway | Kjeller | JEEP2 | 1967 | 2 | 0.22 | 8 | 269 |
| US | Gaithersburg | NBSR/NIST | 1969 | 20 | 2.0 | 17 | 250 |
| France | Grenoble | HFR-ILL | 1972 | 58 | 12.0 | 32 | 225 |
| Germany | Berlin | BER-2 | 1973 | 10 | 2.0 | 16 | 240 |
| France | Saclay | Orphée | 1980 | 14 | 3.0 | 25 | 240 |
| Russia | Moscow | IR-8 | 1981 | 8 | 1.0 | 10 | -- |
| US | Sacramento | MNRC | 1990 | 2 | 0.1 | 5 | 50 |
| Switzerland | Villigen | SINQ | 1996 | 1 | 2.0 | 22 | 250 |
| Korea | Taejon | Hanaro | 1996 | 30 | 2.8 | 6 | 252 |
| Australia | Lucas Heights | OPAL | 2007 | 20 | 1.0 | 15 | 300 |
| *Pulsed neutron sources* | | | | | | | |
| US | Argonne | IPNS | 1980 | 7 | 5 | 13 | 147 |
| Japan | Tsukuba | KENS-KEK | 1980 | 3 kW | 3.0 | 16 | 80 |
| Russia | Dubna | IBR2 | 1984 | 2 | 100 | 13 | 104 |
| UK | Didcot | ISIS | 1985 | 160 kW | 20-100 | 28 | 141 |
| US | Los Alamos | LANSCE | 1985 | 56 | 34 | 7 | 100 |
| US | Bloomington | LENS | 2004 | 6 kW | 0.001 | 3 | -- |
| US | Oak Ridge | SNS | 2006 | 1 | 1.5 | 19 | 240 |
| Japan | Tokai | J-PARC | 2007 | 1 | 0.8 | 1 | -- |
| China | Dongguan | CSNS | 2018 | 0.1 | 0.01 | 18 | Under construction |
| ERIC | Lund | ESS | 2023 | 5 | -- | 15 | Under construction |

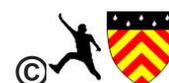

NI offers a typical spatial resolution of a few hundred microns [244] and below 10 micron in the best case [245]. Although XRI is able to provide sub-micron resolution, NI offers better sensitivity to light elements, especially hydrogen [245]. In terms of efficiency, NI may take several hours (days) compared to minutes or even seconds for XRI; this is due to the low neutron flux, dependency on the number of slices/rotation steps and the materials under investigation. The fundamentals, instrumentation and early applications of neutron imaging are covered by Strobl *et al.* [246]; for recent advances and applications refer to Kardjilov *et al.* [247] and Woracek *et al.* [245].

# 4        Conclusions and outlook

NDT techniques are invaluable as tools for testing and evaluation, as may be required during various stages within the lifetime of a composite product. It is clear that each technique has its own potential but rarely achieves the capabilities for a full-scale diagnosis of possible defects and damage evolution in a composite system. Table 9 presents the benefits and limitations of the reviewed NDT methods. Appropriate selection of a suitable NDT technique can be challenging but is clearly essential to provide appropriate information for maintaining the structural integrity of composite materials and structures.

**Table 9        Benefits and limitations of established non-destructive testing techniques used in composite research.**

| NDT technique | Benefits | Limitations | Ref |
|---|---|---|---|
| VI | Simple, rapid, low cost, easy handling | Only for surface flaws or damage<br>Micro-defects are hardly detected<br>Highly subjective, low repeatability and high reproducibility errors<br>Multiple engineering approaches need to be applied for subsurface flaws | [88] |
| AE | Provide real-time monitoring on growing flaws and damage<br>Highly sensitive to stress waves<br>Suitable for *in-situ* and field tests<br>Cover large measurement volumes | Specimen must be stressed<br>Sensitivity is affected by surrounding noise<br>Not suitable for thick specimen<br>Difficult to interpret and characterise damage modes<br>High cost of equipment and consumables<br>High acquisition rates and measurements on test specimen are critical | [3,54,56,57] |

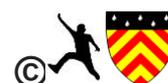

| | | | |
|---|---|---|---|
| UT | Suitable for various material systems<br>Able to detect, locate and size internal flaws<br>Allows one-sided inspection<br>Rapid scan and long-range inspection<br>Good for assembly lines<br>Compact and portable equipment, suitable for on-site inspection<br>Often cost-effective<br>Non-ionizing radiation | Complex setup and transducer design<br>Need skills and experience on multi-modes and complex features<br>Sensitive to operational and environmental surroundings<br>Hard to detect defects near probe<br>Resolution may be limited by algorithms and computing power | [79,85] |
| IRT | Real-time and large-field visual presentation of defects<br>Suitable for wide selection of materials<br>Allows one-sided inspection<br>Safe and easy to operate<br>Cost-effective and productive<br>Non-ionizing radiation | Vulnerable and sensitive for *in-situ* and field tests<br>Limited by cost and availability of excitation sources in field<br>Reduced accuracy with complex geometries<br>Data processing is time-consuming; depends on computing power and algorithms | [100,106,127,248] |
| THz | Robust and repeatable, high scan rate with imaging<br>High precision, sensitivity and resolution<br>High penetration depths for most composites<br>Non-ionizing radiation | Low speed of examination<br>Restricted to nonconductive materials<br>Costly | [2,47,146] |
| ST | Non-contact and full-field surface strain measurement<br>More resilient to environmental disturbance<br>Suitable for large composite structures<br>Efficient for high speed, automated inspection in production environments | External excitation sources are required<br>Limited tolerance to rigid-body motion<br>Limited capabilities for subsurface damage detection<br>Accuracy depends on various sources of uncertainties | [49,158,178] |
| DIC | Cost-effective and easy implementation<br>Adjustable spatial and temporal resolution<br>Insensitive to ambient variations | Speckle patterns are required with high quality<br>Accuracy depends on speckle pattern<br>Limited capability for subsurface detection | [196,197] |
| XRI | Suitable for various materials and *in-situ* tests<br>Can detect both surface and bulk defects<br>2D and 3D images reveal very detailed shape of defects<br>Special resolution at sub-micron level<br>High efficiency<br>Extensive image processing capability | Not suitable for large size structures<br>Not suitable for in-field tests<br>Access to both sides required<br>Dangerous ionizing radiation, requires protection<br>Facilities and access are limited | [212,218,249] |
| NI | Suitable for various materials and *in-situ* tests<br>Can detect both surface and bulk defects<br>2D and 3D images reveal the nature and detailed shape of defects<br>Special resolution at sub-millimetre level<br>Extensive image processing capability<br>Penetration depth greater than X-rays<br>High sensitivity to light elements | Not suitable for in-field tests<br>Access to both sides required<br>Dangerous ionizing radiation, requires protection<br>Acquisition efficiency lower than XRI<br>Facilities and access are limited and more expensive than XRI | [212,245,247] |

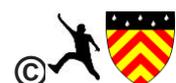

The applications and capabilities of each reviewed NDT technique for detection and evaluation of defects and damage evolution in composite materials/structures are summarised in Table 10. As the volume and structural complexity of composite parts continue to grow, the uses of multi-NDT techniques are becoming increasingly popular in maintaining structural integrity; research in this approach is also growing rapidly.

**Table 10**    **Applications and capabilities of established NDT techniques for composite inspection and evaluation.**

| Defects/damages | AE | UT | IRT | THz | ST | DIC | XRI | NI |
|---|---|---|---|---|---|---|---|---|
| ***Manufacturing induced flaws/defects*** | | | | | | | | |
| Surface defects | | | [250] | [251] | [252] | | [253] | |
| Subsurface defects | [254] | [255, 256] | [250, 257] | [141,145, 251] | [153,258] | | [259] | [260, 261] |
| Delamination | [262] | [263] | [263] | [264] | [265] | [266] | | |
|   Location of depth/size | | [267] | [268] | [90] | [153] | | [267] | |
| Interface debonding | [70] | [269] | [270] | [271] | | | [272] | |
| Fibre content and orientation | | [31] | [273] | [141] | | [274] | [275] | [276] |
| Cracks, fibre breakage/fracture | [277, 278] | [279] | [280] | [141] | [191] | [203] | [281] | [282] |
| Moisture, liquids, or voids | | | [283] | [284] | [154] | | [285] | [286, 287] |
| Shape/surface profiling | | | | | [168] | | | |
| ***In-service/in-situ detection*** | | | | | | | | |
| Impact damage | | [91,288] | [112] | [289] | [91,156,172] | [290] | [290] | |
| Tensile cracking/damage | | [278, 291] | | | | [207] | [292] | [292] |
| Compressive failure | | | [293] | | [293] | | | |
| Bending fatigue damage | [294] | | | [295] | | [294] | [294] | |
| Tensile-tensile fatigue | [296] | | | | | [199] | [297] | [298] |
| Structural damage monitoring | [299] | | [300] | [301] | | [302] | | |
| Quantitative strain monitoring | | | | | [303] | [202] | [304] | [237] |
| Mechanical analysis | | | | | [166,167] | [206] | | |
| Hygrothermal damage | | | | [305] | | | [306] | |

The initial development and application of various NDT techniques are driven by demands from aerospace industries, which rapidly expand to other fields. AE, ultrasound, IRT, shearography, DIC and X-ray imaging represent the main techniques within composite industries, and continue to play essential roles especially in aerospace, automotive, marine and construction applications. NDT techniques based on ultrasound, IRT and DIC are versatile and cost-effective solutions, which have been used extensively in many industrial fields, as well as academic research tools. THz waves can pass through opaque materials and detect internal defects and damage; thus it is a promising NDT technique, possessing great potential in the near future. Innovation and development of compact and portable NDT devices will continue to have a major role for future NDT equipment as these will offer in-service or *in-situ* inspections to facilitate the decision making process.

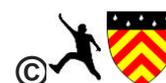

Considering the complex nature of defects and damage detection in composites, the future development of NDT techniques will increasingly depend on intelligent and automated inspection systems with high accuracy and efficiency in data processing. Machine Learning and deep learning provides significant potential for the NDT evaluation of composite materials – artificial intelligence based approaches offer fast decision making without human interference. Various automated diagnostic systems have been proposed for different NDT techniques to offer fast and accurate analysis. These are achieved through artificial neural network coding or algorithms to enable automatic detection and recognition of defects/damage. Examples include: applying pattern recognition to discriminate failure modes in composites using acoustic emission data [307]; damage classification in CFRP laminates using artificial neural networks in ultrasonic testing [308]; automatic defect detection through infrared thermography in CFRP laminates [309] and honeycomb composite structures [310]; an automated shearography system for cylindrical surface inspection using machining learning [252]; neural network-based hybrid signal processing for terahertz pulsed imaging [311]. Despite the exciting achievements in NDT techniques, there is still substantial work required to develop fast and affordable systems for both equipment and data processing methods to promote their practical implementation in industry.

Although X-ray and neutron imaging are powerful tools for NDT tests offering super high resolution, both imaging techniques are based on ionizing radiation; *i.e.* specimens have to be analysed using radiation facilities which are inconvenient compared to other NDT techniques. Also, locations and availability of synchrotron facilities are very limited, which further constrain their accessibility and costs. Portable X-ray or neutron generators have been commercialised to provide easier access. Whilst they have found applications in aerospace, marine, construction and pipeline inspection, their capabilities for composite industries are limited. The use of neutron imaging depends on advances in neutron production and instrumentation, whilst its research community is growing rapidly. Free electron lasers and modern spallation sources are promising techniques that should enhance the future development of NDT technology towards more advanced capabilities.

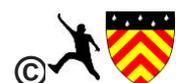

# Funding

The author(s) received no financial support for the research, authorship, and/or publication of this article.

# Declaration of conflicting interests

The author(s) declared no potential conflicts of interest with respect to the research, authorship, and/or publication of this article.

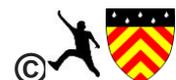

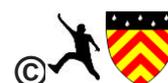

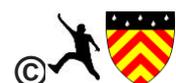

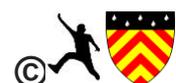

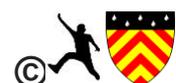

and flat panel composite structures used in aerospace applications - E2661 2015:1–19.

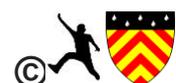

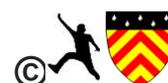

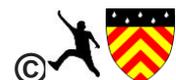

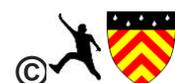

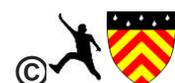

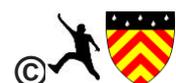

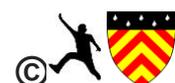

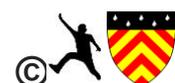

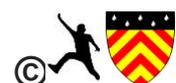

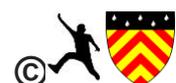

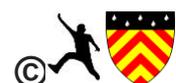

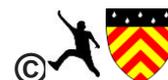

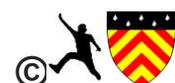

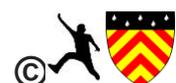

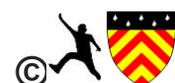

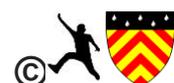

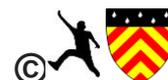

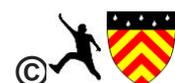